\documentclass{article}
\usepackage{ltwol2e}
\usepackage{epsf}

\newcommand{\wt}{\widetilde}
\newcommand{\NN}{{\cal N}}
\newcommand{\RR}{{\cal R}}
\newcommand{\refb}[1]{(\ref{#1})}
\def\be{\begin{equation}}
\def\ee{\end{equation}}
\def\bea{\begin{eqnarray}}
\def\eea{\end{eqnarray}}
\begin{document}

\title{DEVELOPMENTS IN SUPERSTRING THEORY} 

\author{Ashoke Sen}

\address{Mehta Research Institute of Mathematics and Mathematical
Physics, \\
Chhatnag Road, Jhusi, 
Allahabad 211019, INDIA\\E-mail: sen@mri.ernet.in}


\twocolumn[\maketitle\abstracts{ 
In this talk I review recent
developments in superstring theory. In the first half of the talk
I discuss some of the earlier developments in this subject.
This includes a
review of the status of string theory as a unified theory of
gravity and other interactions, the role of duality in string
theory, and application of string theory to the problem of
information loss near a black hole. In the
second half of the talk I review the developments of the
last two years. This includes application of duality symmetries,
Matrix theory, Maldacena
conjecture, and derivation of gauge theory results from string
theory. In choosing the list of topics for this talk, I have
focussed on those which are likely to be of some interest to
non-string theorists, thereby leaving out many of the marvelous
technical results in this subject.}]

\section{Earlier Developments}

\subsection{String Theory and Quantum Gravity} \label{s11}

\begin{figure}[!ht]
\begin{center}
\leavevmode
\epsfbox{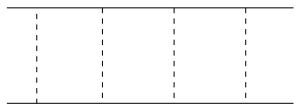}
\end{center}
\caption{Divergent contribution to $e^--e^-$ scattering in
quantum general relativity. In this diagram the solid lines
denote the electron propagator and the broken lines denote
graviton propagator.}
\label{f1}
\end{figure}

As we all know, quantum field theory has been extremely
successful in providing a description of elementary particles and
their interactions. However, it does not work so well for
gravity. If we naively try to quantize general relativity $-$
which is a classical field theory $-$ using the methods of quantum
field theory, we run into divergences which cannot be removed by
using the conventional renormalization techniques of quantum
field theory. An example of such a divergent graph contributing to
$e^--e^-$ scattering has been shown in Fig.\ref{f1}.

\begin{figure}[!ht]
\begin{center}
\leavevmode
\epsfbox{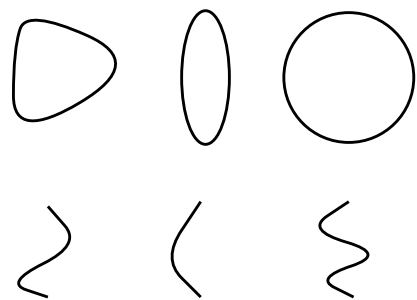}
\end{center}
\caption{Some vibrational modes of closed and open strings}
\label{f2}
\end{figure}

String theory is an attempt to solve this problem.\cite{GSW}
The basic idea in string theory is quite simple. According to
string theory, different elementary particles, instead of being
point like objects, are different vibrational modes of a string.
Fig.\ref{f2} shows some of the oscillation modes of closed strings
and open strings. However, the typical size of a string is
extremely small, being of the order of the Planck length ($\sim
10^{-33}cm.)$. Thus in all present day experiments these will
appear to be point-like objects, and string theory will be
indistinguishable from an ordinary quantum field theory.

This simple idea has drastic consequences. One finds
that 
\begin{itemize}
\item A consistent quantum string theory does not suffer from any
ultraviolet divergences.
\item The spectrum of such a string theory automatically contains a
massless spin two state which has all the properties of a
graviton $-$ the mediator of gravitational interaction.
\end{itemize}
Thus string theory automatically gives us a finite quantum theory
of gravity! However, there are problems in the immediate application
of string theory to our world. One finds that:
\begin{itemize}
\item
String theory is consistent only in (9+1) dimensional space-time
instead of the (3+1) dimensional space-time in which we live.
\item Instead of a single string theory there are five consistent
string theories in (9+1) dimensions. They are named as:
\begin{center}
Type IIA, \quad Type IIB, \quad Type I, 

$E_8\times E_8$
heterotic \quad and \quad SO(32) heterotic
\end{center}
\end{itemize}

As we shall see, the first problem is resolved using the idea of
compactification. The second problem is partially resolved using
the idea of duality.

\subsection{Compactification} \label{s12}

\begin{figure}[!ht]
\begin{center}
\leavevmode
\epsfbox{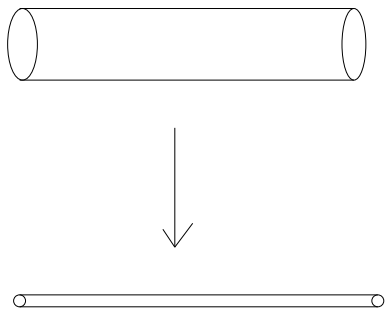}
\end{center}
\caption{Example of Compactification. The two dimensional
cylinder appears to be one dimensional if the radius is very
small.}
\label{f3}
\end{figure}

The idea of compactification is quite simple. Although
consistency of string theory tells us that it must live in (9+1)
dimensions, there is no need for all the 9 space-like directions
to be of infinite extent. Instead we can take six of them to be
small and compact, and the other three to be of infinite extent.
As long as the sizes of the compact directions are smaller than
the resolution of the most powerful microscope (the
accelerators), we shall not be able to see these extra directions
and the world will look (3+1) dimensional. This has been
explained in Fig.\ref{f3} by a toy example. Here we have an
intrinsically two dimensional surface, but one of the two
directions is taken to be a circle of small radius. For
sufficiently small radius of the circle, this two dimensional
cylinder appears to be a one dimensional space.

It turns out that there are many different choices for this six
dimensional compact manifold. Thus each of the five string
theories in (9+1) dimensions gives rise to many possible string
theories in (3+1) dimensions after compactification. Some of
these theories come tantalizingly close to the observed universe.
In particular one can construct models with:
\begin{itemize}
\item Gauge group containing the standard model gauge group
$SU(3)\times SU(2)\times U(1)$,
\item Chiral fermions representing
three generations of quarks and leptons,
\item N=1 supersymmetry,
\item Gravity.
\end{itemize}
Furthermore unlike conventional quantum field theories which are
ultraviolet divergent but renormalizable, and quantum general
relativity which is ultraviolet divergent and not renormalizable,
string theories have no ultraviolet divergence at all!

\subsection{Duality Symmetries of String Theory}

Existence of duality symmetries in string theory started out as a
conjecture and still remains a conjecture. However so many
non-trivial tests of these conjectures have been performed by now
that most people in the field are convinced of the validity of
these conjectures. A review of this subject and more references
can be found in ref.\cite{AS}.

A duality conjecture is a statement of equivalence between two or
more apparently different string theories. Two of the most
important features of duality are as follows:
\begin{itemize}
\item Often
under the duality map, an elementary particle in one theory
gets mapped to a composite particle in a dual theory and vice
versa. Thus classification of particles into elementary and
composite loses significance as it depends on which particular
theory we use to describe the system.
\item Often duality relates a weakly coupled string theory to a
strongly coupled string theory and vice versa. In many simple
cases the coupling constants $g$ and $\wt g$ in the two theories 
are related via the simple relation:
\be \label{e1}
g = \wt g^{-1}\, .
\ee
Thus a perturbation expansion in $g$ contains information about
non-perturbative effects in the dual theory. In particular the
tree level (classical) results in one theory can contain
contribution from perturbative and non-perturbative terms in the
dual theory. This also clearly shows that duality is a property
of the full quantum string theory, and not of its classical
limit.
\end{itemize}
Thus there are two aspects of duality
\begin{center}
elementary $\leftrightarrow$ composite

classical $\leftrightarrow$ quantum
\end{center}

Let me now give some examples of dual pairs of string theories.
\begin{itemize}
\item (9+1) dimensional
SO(32) heterotic and type I string theories are conjectured
to be dual to each other.\cite{WITTEND,POLWIT}
\item SO(32) heterotic string theory compactified on a four
dimensional torus (denoted as $T^4$)
is conjectured to be dual to type IIA
string theory compactified on a different four dimensional
manifold, denoted by $K3$.\cite{HT}
\item Type IIB string theory is conjectured to be self-dual, in
the sense that the type IIB string theories at two different
couplings $g$ and $\wt g$ related by eq.\refb{e1} are conjectured
to describe the same physical theory.\cite{HT}
\item Heterotic string theory compactified on a six dimensional
torus, denoted by $T^6$, is conjectured to be self-dual in the same
sense as above.\cite{FONT,SREV}
\end{itemize}

\begin{figure}[!ht]
\begin{center}
\leavevmode
\epsfbox{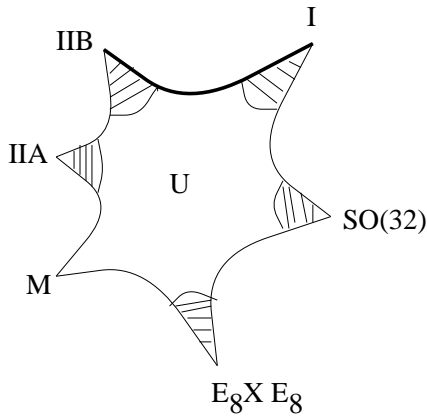}
\end{center}
\caption{Moduli space of unified string theory. The shaded
regions denote weakly coupled string theories, and can be studied
using perturbation theory. Most of the region in the U-theory
moduli space, depicted in white, does not admit such a
description.
}
\label{f4}
\end{figure}
Due to the fact that a duality conjecture relates two 
apparently different theories, we see that it gives a unified
picture of all string theories. The situation is summarized in
Fig.\ref{f4}.\footnote{
One should keep in mind that this is only a
schematic representation. Each string theory gives rise to
many weakly coupled string theories after compactification. Also
different compactifications of a string theory may be separated by
infinite distance in the moduli space if in order to go from one
to the other one needs to pass through the decompactification
limit. Finally, not all components of $U$-theory may belong to
the same moduli space, {\it i.e.} there may not be a continuous
path connecting them such that each point on the path represents
a solution of the field equations~\cite{DABHAR}, 
but it is expected that they
are all connected via intermediate configurations which are not
necessarily solutions of field equations.
} According to this picture the apparently
different string theories and their compactifications
are just different limits of the same theory, with a large
parameter space.\footnote{In string theory parameters themselves
are related to vacuum expectation values of different fields and
are expected to be determined dynamically.} There is no
universally accepted name for this central theory, $-$ I have
chosen to call it $U$-theory for the purpose of this talk. 
$U$ can be
taken to stand for Unknown or Unified. Some small
regions of the parameter space of
$U$-theory, which can be represented by some {\it weakly
coupled} string theory, are reasonably well understood. This has
been denoted by the shaded regions in Fig.\ref{f4}, 
and correspond to the weak
coupling regime of the five different string theories and their
compactifications. But for
most of the parameter space $U$-theory does not have a description
in terms of weakly coupled string theory. Note that in one corner
of the parameter space of $U$-theory, there
is a theory called $M$-theory~\cite{TOWN,WITTEND,SCHM,ASPINM,HORWIT} 
which has not been introduced before.
At present not much is known about $M$-theory except that its low
energy limit is the eleven dimensional supergravity theory, 
and that various
string theories and their compactifications
approach $M$-theory in certain limits. However, unlike
string theory, $M$-theory does not have any coupling constant,
and no systematic procedure for doing computations in 
$M$-theory beyond the low energy supergravity limit is known. (As
we shall discuss later, Matrix theory is an attempt in this
direction.) It is
generally believed that due to the absence of a coupling
constant, understanding $M$-theory will require a full
understanding of $U$-theory, and for this reason $U$-theory has often
been identified with $M$-theory. However, we shall keep the
distinction between the two, and reserve the name $M$-theory for
the unknown eleven dimensional theory which arises in a certain
limit of $U$-theory.

Finally we note that since duality typically relates a strongly
coupled string theory to a weakly coupled string theory, and
since at present there is no independent description of string
theory at strong coupling, we cannot prove duality. However one
can device various tests of duality by working out various
consequences of a duality conjecture, and then verifying these
predictions by explicit computation. Often such predictions take
the form of highly non-trivial mathematical identities which can
nevertheless be proved. (See refs.\cite{ASDUAL,VWIT,KAPL} for 
specific examples.)
As I have already stated, as a result of
many such tests, most people in the field now take it for
granted that duality is a genuine symmetry of string theory.

\subsection{String Theory and the Information Loss Puzzle}

Black holes are classical solutions of general relativity, and
its extensions like string theory. They can be formed by collapse
of matter under its own gravitational pull, or can be primordial,
formed at the time of the big bang. Classically black holes are
completely black, in the sense that they absorb everything that
falls within a certain radius (known as the event horizon) and
nothing can come out of a black hole once it is inside the event
horizon. However this picture changes dramatically in quantum
theory. It has been argued by Hawking, Bekenstein and
others~\cite{HAWK,BEKEN} that
in quantum theory a black hole behaves as a perfect black body at
finite temperature (inversely proportional to its mass). In
particular,
\begin{itemize}
\item Black holes emit thermal black body radiation.
\item Black holes carry entropy proportional to the area of its
event horizon.
\end{itemize} 
This thermal description of the black hole can cause potential
conflict with quantum mechanics and give rise to the information
loss puzzle. For this let us consider the following thought
experiment.
\begin{itemize}
\item we start with a pure quantum state in the $s$-wave and let
is collapse under its own gravitational pull to form a black
hole.
\item We then wait till it evaporates via Hawking radiation.
\end{itemize}
Since the Hawking radiation is thermal, the final state
represents a mixed state. Thus the net result of this process is
the evolution of a pure quantum state into a mixed state. In
particular, most of the information about the initial state
(except its mass and gauge charges) are lost during the process.
This is certainly in conflict with the fundamental principles of
quantum mechanics.

In order to put things in perspective, let us note that all hot
objects emit thermal radiation, but they do not violate the
principles of quantum mechanics. The main (apparent) difference
from the case of the black hole is that for an ordinary hot
object the thermal description is merely a matter of convenience,
and represents our inability to explicitly study the quantum
mechanics of large number of particles. Thus we {\it choose} to
give a statistical description of the object by averaging over
microstates. However
we can, in principle, give a completely microscopic
description of the radiation in terms of quantum scattering
processes occuring inside the object.

The absence of such a microscopic description in
case of thermal radiation from a black hole is the main cause of
the information loss puzzle. Thus if one could provide a
microscopic description of the Bekenstein
entropy and Hawking radiation of a black hole in 
terms of quantum states of the hole and their scattering 
respectively, then the thermal radiation from the black hole will
be on the same footing as the radiation from any other hot
object; and the information loss problem will disappear. This is
precisely what string theorists have attempted to do and have
been partially successsful. It was found that for a special class
of black holes:
\begin{itemize}
\item One can count the number of quantum states $N$ of these
black holes and show that:~\cite{SUSSBH,ASBH,STRVAF}
\be \label{e2}
S_{BH} = \ln(N)\, ,
\ee
where $S_{BH}$ is the Bekenstein-Hawking entropy of a black hole
and $N$ is the degeneracy of states of the black hole carrying
certain quantum numbers. This gives a microscopic explanation of
the Bekenstein-Hawking entropy of the black hole.
\item One can compute the rate of Hawking radiation from these
black holes due to quantum scattering processes inside the hole
and show that the rate agrees precisely with the Hawking
radiation formula.\cite{CALMAL,MDW,DASMAT,STRMAL} 
Thus this provides a microscopic explanation of Hawking
radiation.\footnote{In both these computations the microscopic
computation is done in a region of the parameter space where
gravity is weak, and then continued to the regime of strong
gravity with the help of a non-renormalization theorem.}
\end{itemize}
Although these calculations have been done for a special class of
black holes for special regions in the parameter space, at least
this gives us hope that in string theory, the phenomenon of
Hawking radiation and the entropy of a black hole are firmly
embedded within the usual principles of quantum mechanics.

\section{Recent Developments}

Much of the recent development in the subject have focussed on
the study of the strong coupling regime of string theory, which
might also be described as the study of $U$-theory. We shall
describe some of these approaches here.

\subsection{Application of Duality to Deriving U-theory Effective
Action}

This approach is best understood with the help of Fig.\ref{f4}.
As we have said earlier, the shaded regions in this diagram
denote various weakly coupled string theories, and hence various
quantities of interest can be computed in these regions using
string perturbation theory. Given these results, one can try to
guess the exact form of the answer for these quantities by
demanding that it correctly interpolates between the known
results in various shaded regions, and is consistent with various
duality symmetries and known perturbative non-renormalization
theorems. This approach has yielded
several concrete results in $U$-theory. Thus for example, this
has been used to deduce the correct form of some specific terms in
the effective action of type IIB string
theory compactified on tori of various 
dimensions.\cite{GREE,BERVAF,KIRPIO} Here we quote one such term
in the effective action of type IIB string theory:
\be \label{exyz}
{1\over 3\cdot 2^{14}\cdot \pi^7\cdot\alpha'} \int d^{10}x
\sqrt{-det g} \, E(\tau,\bar\tau) \RR^4\, ,
\ee
where $g_{\mu\nu}$ denotes the metric, $\RR^4$ is a specific
scalar constructed from the fourth power of the Riemann curvature
tensor, $\tau$ is a complex scalar field, 
$\alpha'$ is related to the string tension $T_S$ as
$T_S=(2\pi\alpha')^{-1}$, and
\be \label{exyz1}
E(\tau,\bar \tau) = (Im\tau)^{3\over 2} {\sum_{m,n}}^\prime 
{1\over |m+n\tau|^2}\, .
\ee
$\Sigma'$ denotes sum over all integers $m$, $n$ except for
$m=n=0$.
These terms
contain both perturbative and non-perturbative contributions from
the viewpoint of type IIB string theory, and whereas the
perturbative terms could be explicitly derived using string
perturbation theory, the non-perturbative terms are novel
results which are beyond the reach of string perturbation theory.

Of all the attempts to understand $U$-theory, this approach has
been the most successful one in deriving concrete new
quantitative results in
$U$-theory. Unfortunately this approach probably can be used to
derive only a limited set of terms in the $U$-theory effective
action $-$ terms which enjoy some sort of perturbative 
non-renormalization
theorem $-$ since such guesswork becomes quite difficult
as one considers more general terms in the effective action
of $U$-theory.

\subsection{Matrix Theory}

Going back to Fig.\ref{f4}, we notice that in one corner of the
moduli space of $U$-theory is a theory called $M$-theory living
in eleven space-time dimensions. We have
said earlier that not much is known about this theory, except
that its low energy limit is the eleven dimensional supergravity
theory. This theory does not contain any freely adjustable
coupling constant, and hence there is no systematic perturbation
expansion which allows us to compute amplitudes beyond the low
energy supergravity approximation. Matrix theory is an attempt to
give a nonperturbative definition of $M$-theory.\cite{BFSS}
Since most of the
known regions of $U$-theory can be regarded as some sort of
compactification of $M$-theory, there is also a hope that by
giving a non-perturbative definition of $M$-theory and its
various compactifications~\cite{WT,SUSS,ASMAT,SEIMAT} one might be
able to give a description of the full $U$-theory.

In order to understand the basic proposal of Matrix theory, let
us consider a specific scattering process in $M$-theory. Instead
of trying to compute the amplitude in the center of momentum
frame, we consider a frame which is boosted by a large (infinite)
amount relative to the center of momentum frame in some given
direction. This frame is sometime known as the infinite momentum
frame or the light-cone frame, as all external particles
participating in the scattering process travels with infinite
momentum along the forward light-cone. According to the Matrix
theory proposal {\it $M$-theory in the infinite momentum frame is
equivalent to a quantum mechanical system in the sense that
computation of every scattering amplitude in $M$-theory can be
mapped to the computation of an appropriate correlation function
in this quantum mechanical system.   The fundamental
degrees of freedom of this quantum mechanical system
are $N\times N$ matrices, the Hamiltonian is that of a
particular supersymmetric quantum mechanics, and we need
to take the limit $N\to\infty$ at the end of the calculation.}

Since the Hamiltonian of this quantum mechanical system is given,
this in principle gives an algorithm to compute any amplitude in
$M$-theory, and thereby provides a non-perturbative definition of
$M$-theory. The most obvious consistency check that this proposal
can be subjected to is that in the low energy limit, the
amplitudes computed from the matrix theory proposal should agree
with those computed from eleven dimensional supergravity. This
has been tested in many examples.\cite{BECBEC,BBPT,VERVER,OKAYON}
Unfortunately, the complexity
of matrix quantum mechanics has so far prevented us from deriving
any new quantitative result about $M$-theory using Matrix theory.
In particular, the large $N$ limit of this quantum mechanical
system is not well understood. Although it has not affected the
computations which have been done so far, 
it has been suggested 
that the full agreement between the supergravity and matrix theory
computations may involve subtle effects of the large N limit when we
consider more complicated amplitudes.\cite{DOS,KABPOL,DDM,DINE}
For these reasons, Matrix theory is still in its infancy. 
It would indeed be interesting to see if Matrix theory can be
used to derive some of the terms in $U$-theory, which have been
derived using the approach outlined in the previous section. 

\subsection{Maldacena Conjecture}

A third approach to the study of $U$-theory is based on the 
discovery of a new relationship between some region of $U$-theory, 
and ordinary quantum field theories. This series of conjectures
was first put forward by Maldacena,\cite{MALD} and hence 
is now known as the
Maldacena conjecture. In this talk I shall focus on only one of
these conjectures. It states that:

\noindent{\it Type IIB string theory on $S^5\times AdS_5$ is
equivalent to $\NN=4$ supersymmetric $SU(N)$ gauge theory in (3+1)
dimensions.}

Since this conjecture involves several terms which might not be
familiar, let me try to define each of these terms. First of all
$S^5$ denotes a five dimensional sphere, which might be described
by its embedding in six dimensional Euclidean space with
coordinates $y^1,\ldots y^6$ as follows:
\be \label{e3}
\sum_{i=1}^6 (y^i)^2 = R^2\, .
\ee
$R$ denotes the radius of the sphere. $AdS_5$, $-$ the five
dimensional anti- deSitter space $-$ can be described by its
embedding in a six dimensional flat space with signature
($--++++)$ spanned by coordinates $x^0,\ldots x^5$ as follows:
\be \label{e4}
(x^0)^2 + (x^1)^2 -\sum_{i=2}^5 (x^i)^2 = R^2\, .
\ee
$R$ is a parameter labelling the size of the $AdS_5$ space. Note
that we have used the same parameters $R$ for both $S^5$ and
$AdS_5$. This is intentional, since the $S^5$ and $AdS_5$
appearing in the Maldacena conjecture are related this
way.\footnote{Many of the concrete results arising out of this
conjecture are better understood in its Euclidean version,
obtained by making a Wick rotation $x^0\to i x^0$ in
eq.\refb{e4}. On the gauge theory side we consider $\NN=4$
supersymmetric $SU(N)$ gauge theory in the
four dimensional Euclidean
space instead of the (3+1) dimensional Minkowski space.}

Thus $S^5\times AdS_5$ is a ten dimensional space, and one side
of the duality relation involves type IIB string theory
on this ten dimensional space.\footnote{There are
other background fields present on this space, but we shall not
describe them here.} Let us now turn to the other side of the
duality conjecture. An $\NN=4$ supersymmetric $SU(N)$ gauge theory
is an ordinary $SU(N)$ gauge theory with four Majorana fermions and
six scalars in the adjoint representation of the gauge group.
Furthermore, all the Yukawa couplings and the scalar
self-couplings are completely determined in terms of the gauge
coupling constant using the requirement of
supersymmetry.\cite{OSBORN} Thus
the only free parameters in the theory are the gauge coupling
constant, and the vacuum angle $\theta$ which does not affect
perturbation theory.

This finishes the definition of all the terms that appeared in
the Maldacena conjecture. Since the conjecture relates a ten
dimensional theory to a four dimensional theory, the exact
meaning of this conjecture is not transparent. The precise form
of this conjecture states that {\it there is a one to one
correspondence between the physical green's functions in type IIB
string theory on $S^5\times AdS_5$ and the correlation functions
of gauge invariant operators in $\NN=4$ supersymmetric $SU(N)$
gauge theory.}\cite{WITONE,GUPOKL} 
Such a relationship is possible because the
physical excitations in $AdS_5$ live on the boundary of $AdS_5$,
which in turn can be identified with a four dimensional flat
space-time (after including `points at infinity').

Type IIB string theory on $S^5\times AdS_5$ has three
dimensionless parameters, $-$ the string coupling constant
$g_{st}$, a constant $a$ related to the vacuum expectation
value of a particular scalar field in the theory (known as the
Ramond-Ramond scalar), and the ratio
$R/\sqrt{\alpha'}$,
where 
$R$ is the common radius of $S^5$ and
$AdS_5$. On the other hand $\NN=4$ supersymmetric Yang-Mills
theory is also characterized by three parameters, the gauge
coupling constant $g_{YM}$, the vacuum angle $\theta$ and the
number $N$ determining the gauge group $SU(N)$. The Maldacena
conjecture gives a precise relation between these parameters:
\be \label{e5}
g_{st}=g_{YM}^2, \qquad a = \theta, \qquad {R\over \sqrt{\alpha'}}
= (4\pi g_{YM}^2 N)^{1/4}\, .
\ee

There are many other examples of such conjectures which have been
found. Typically such a conjecture takes the form of an
equivalence relation between a string theory / $M$-theory on a certain
manifold $K$, and a quantum field theory. The precise form of this
quantum field theory depends on the choice of the manifold $K$ as
well as on which theory we compactify. Since quantum field
theories have intrinsic non-perturbative definition (even if
explicit computations may be difficult) we can use this
conjecture to give non-perturbative definitions of
string theories in specific backgrounds. This makes certain regions of
$U$-theory accessible via the usual quantum field theory
techniques. Here we shall discuss one particular application of
the Maldacena conjecture to the study of $U$-theory $-$ namely
the derivation of the holographic principle.

The holographic principle, first proposed by 't Hooft and then by
Susskind,\cite{HOOF,SUSSHOL} states that {\it in a 
consistent quantum theory of gravity,
the fundamental degrees of freedom reside at the boundary of
space-time and not in the interior. Furthermore, on the boundary
there is precisely one degree of freedom per Plank area.} (If the
original theory is $(d+1)$-dimensional, then here Planck `area'
refers to a $(d-1)$ dimensional
volume of order $M_{pl}^{-d+1}$, where $M_{pl}$ is
the Planck mass.) The Maldacena conjecture provides a concrete
verification of this holographic principle for type IIB on
$S^5\times AdS_5$. According to this conjecture the degrees of
freedom in this theory can be identified as those of an $\NN=4$
supersymmetric $SU(N)$ gauge theory living on the boundary of
$AdS_5$. After proper ultraviolet regularization of the gauge
theory (which amounts to infrared regularization in the type IIB
string theory) one can count the total number of degrees of
freedom of the gauge theory. Dividing this by the volume of the
boundary one finds of order one degree of freedom per
Planck volume, as required by the holographic
principle.\cite{SUSWIT}

Although this illustrates the holographic principle in $AdS$
space, extension of this principle to Minkowski space remains an
open and challenging problem.

\section{Application of String Theory to Gauge Theories}

We shall now discuss application of string theory to gauge
theories. Since we have been discussing Maldacena conjecture, we
shall first discuss its application, and then turn to the
(historically older) subject of gauge theories from branes.

\subsection{Maldacena Conjecture}

Since the Maldacena conjecture relates string theory to gauge
theory, we can use it to study gauge theories using known
results in string theory. This may sound strange, as we
understand gauge theories much better than string theory; however
as we shall see, in some region in the parameter space the string
theory side is better understood, and hence can be useful in
deriving gauge theory results. For this let us look at the
relation \refb{e5} between the string theory and the gauge theory
parameters. If we consider the 't Hooft large $N$
limit~\cite{HOOFN} on the gauge theory side:
\be \label{e6}
g_{YM}\to 0, \qquad N\to\infty, \qquad \lambda\equiv g_{YM}^2N
=\hbox{fixed}\, ,
\ee
then it corresponds to the following limit in string theory:
\be \label{e7}
g_{st}\to 0, \qquad {R\over \sqrt{\alpha'}}= (4\pi \lambda)^{1/4}
=\hbox{fixed}\, .
\ee
The smallness of $g_{st}$ implies that in this limit we can use
classical string theory.
Let us now further take the limit when the 't Hooft coupling
$\lambda$ is large. In this case $R/\sqrt{\alpha'}$ is large and
hence we can approximate string theory by its supergravity limit
for computation of Green's functions with external momenta small
compared to the string scale. Since these Green's functions are
related to the correlation functions of gauge invariant operators
in the gauge theory, we conclude that in the t' Hooft large $N$
limit with {\it large 't Hooft coupling $\lambda$}, the gauge theory
correlation functions can be computed by studying classical type
IIB supergravity theory on $S^5\times AdS_5$! Note that ordinary
large $N$ perturbation theory in gauge theory is useless in this 
limit, as it involves a series expansion in $\lambda$. Thus this
approach can give genuinely new results in gauge theory.

\begin{figure}[!ht]
\begin{center}
\leavevmode
\epsfbox{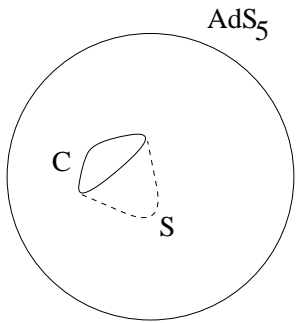}
\end{center}
\caption{Computation of Wilson line $-$ or equivalently, the quark
anti-quark potential $-$ using Maldacena conjecture. Here $C$ is a
closed curve in the four dimensional space-time, regarded as the
boundary of $AdS_5$, and $S$ is a surface of minimal area
in the interior of $AdS_5$ subject to the condition that $C$ 
is its boundary. The Wilson line associated with the curve $C$ in
the $\NN=4$ supersymmetric gauge theory is given by the area of
the surface $S$ in $AdS_5$.}
\label{f5}
\end{figure}
Here I shall discuss one example $-$ the computation of
(non-dynamical) quark anti-quark potential in the $\NN=4$
supersymmetric gauge theory.\cite{MALWIL,REYEE} It turns out 
that using the
Maldacena conjecture this problem can be mapped to the geometric
problem of finding the minimal area surface bounded by a fixed
curve at the boundary of $AdS_5$ as illustrated in Fig.\ref{f5}. 
This problem is easily solved,
and the final answer for the potential is:
\be \label{e8}
V(r) = -{4\pi^2 (2 g_{YM}^2 N)^{1\over 2}\over (\Gamma({1\over
4}))^4 r}\, ,
\ee
where $r$ is the separation between the quark and the anti-quark.
Although the $r$ dependence of the potential follows from the
conformal invariance of the gauge theory, the dependence on the
coupling constant $g_{YM}$ is novel since it is linear instead of
quadratic as expected from perturbation theory.

One can also use the Maldacena conjecture to study $\NN=4$
supersymmetric gauge theories at finite temperature
$T$.\cite{WITONE,WITTWO} In this case
we need to make the time direction
\begin{itemize}
\item Euclidean, and
\item periodic with period $2\pi T^{-1}$.
\end{itemize}
This theory turns out to be dual to type IIB string theory on
$S^5\times K$, where $K$ is a new five dimensional manifold
representing the Euclidean black hole solution in $AdS_5$.
The boundary of this new manifold $K$ is $R^3\times S^1$, which
is to be identified with the three space and the periodic time
direction of the gauge theory. As in the case of the zero
temperature theory, in the 't Hooft large $N$ limit,
with large 't Hooft coupling $\lambda$ for the gauge theory, the
relevant type IIB string theory can be approximated by classical
type IIB supergravity theory on $S^5\times K$. This description
can be used to study various properties of the finite temperature
gauge theory in this limit. Thus for example, the problem of
finding the mass spectrum in the finite temperature gauge theory 
can be mapped to the problem of finding eigenvalues of certain
differential operators on $K$. As an example we quote the
relevant differential equation for determining the mass $m_n$ of the
$n$th scalar
`glueball' (from the point of view of three dimensional gauge
theory obtained in the high temperature limit of the four dimensional
gauge theory)
\be \label{e9}
\pi^2 T^2 {1\over \rho} {d\over d\rho}\Big[ (\rho^5-\rho)
{df_n\over d\rho}\Big] = -m_n^2 f_n\, ,
\ee
where $\rho$, denoting the `radial coordinate' in $K$, lies in
the range $1\le \rho<\infty$. $f_n$ satisfies the boundary
conditions
\begin{eqnarray} \label{e9aa}
{df_n\over d\rho} &=& 0 \qquad \hbox{at} \quad \rho=1 \nonumber
\\
f_n &\sim& \rho^{-1} \qquad \hbox{as} \quad \rho\to\infty\, .
\end{eqnarray}
This converts a non-perturbative
quantum field theory problem to a classical eigenvalue problem.
This eigenvalue problem can be solved by numerical
methods.\cite{NUMERIC}

So far we have talked about supersymmetric gauge theories. It is
natural to ask if these techniques can be used for getting
information about pure $SU(N)$ gauge theories. It was shown by
Witten~\cite{WITTWO} that pure $SU(N)$ gauge theory 
is dual to $M$-theory on
$S^4\times K_7$, where $K_7$ is a particular seven dimensional
manifold related to the Euclidean black hole solution in the
seven dimentional anti-de Sitter space. The temperature of this
black hole, which is a parameter labelling the manifold $K_7$,
corresponds to the ultraviolet cut-off in the $SU(N)$ gauge theory.
In the limit
\be \label{e10}
N\to\infty, \qquad g_{YM}\to 0, \qquad g_{YM}^2N=\hbox{fixed but
large}\, ,
\ee
$M$-theory can be approximated by classical eleven dimensional
supergravity theory. Thus various properties of gauge theory in
this limit can be studied using the classical supergravity theory
on $S^4\times K_7$. In particular one can prove confinement and
existence of a mass gap in the gauge theory in this limit.
However, since pure $SU(N)$ gauge theory is asymptotically free, in
order to take the continuum limit, one needs to take the
ultraviolet cut-off to infinity, and the 't Hooft coupling
$g_{YM}^2N$ to {\it zero} keeping a certain combination fixed.
Unfortunately in this limit the classical supergravity is no
longer a good approximation to $M$-theory. Thus application of
these ideas to the study of large $N$ gauge theory in the
continuum limit remains an open problem.

\subsection{Gauge Theories from Branes}

Let me now discuss another approach that has been useful in
deriving gauge theory results from string
theory.\cite{FTH,BADOSE,HANWIT,WITBRANE} This involves
the study of branes.
Historically this preceeds the Maldacena conjecture, and in fact
Maldacena arrived at his conjecture by examining a special
configuration of branes.

Branes are static classical solutions in string theory which are 
known to exist in
many string theories. A $p$-brane denotes a static
configuration which
extends along $p$ spatial direction (the tangential directions) and 
is localized in all other
spatial directions (the transverse directions). 
Thus the solution is invariant under
translation along the $p$ directions tangential to the brane, as
well as the time direction, and
approaches the vacuum configuration as we go away from the brane
in any one of the transverse direction. 
Thus in this language, 
\begin{center}
1-brane \qquad $\equiv$ \qquad string

2-brane \qquad $\equiv$ \qquad membrane 

0-brane \qquad $\equiv$ \qquad particle
\end{center}
etc.
Typically the quantum
dynamics of a configuration of $p$-branes is described by a
$(p+1)$ dimensional gauge field theory,\cite{POLDBR,WITDBR} and 
the coupling
constant of this quantum field theory is related to the coupling
constant of the string theory of which the brane configuration is
a solution. In this case duality symmetries relating strong and
weak coupling limits of the original string theory can be used to
derive duality relations involving the quantum field theories
describing the dynamics of the brane. This approach has been used
to derive many different results in supersymmetric gauge
theories. Some example are:
\begin{itemize}
\item Derivation of Montonen-Olive duality~\cite{MONOL} in $\NN=4$
supersymmetric gauge theories.\cite{NEFOUR}
\item Derivation of Seiberg-Witten like results~\cite{SEIWIT} 
in $\NN=2$
supersymmetric gauge theories.\cite{FTH,BADOSE,WITBRANE}

\item
Derivation of a special kind of symmetry, known as the mirror
symmetry,\cite{MIRROR} 
in (2+1) dimensional gauge theories.\cite{HANWIT}

\item
Derivation of Seiberg dualities~\cite{SEIBERG} involving 
$\NN=1$ supersymmetric
gauge theories in (3+1) dimensions.\cite{NEQONE,WITNEONE}
\end{itemize}

\begin{figure}[!ht]
\begin{center}
\leavevmode
\epsfbox{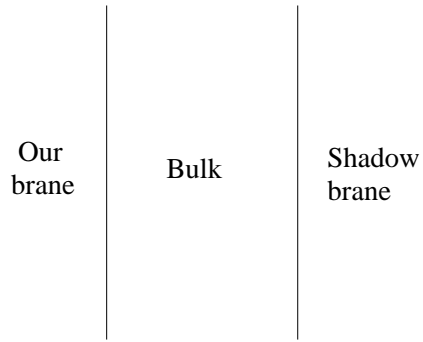}
\end{center}
\caption{Possibility of novel compactifications involving branes.
Our brane hosts the standard model fields, the (9+1) dimensional
`bulk' hosts gravity and its superpartners, and the shadow brane
hosts other fields which communicate to us via gravitational
interaction. Supersymmetry may be broken at a high scale in the
shadow brane, and yet have relatively small effect in our brane
due to the weakness of gravitational interactions.}
\label{f6}
\end{figure}
Existence of branes in string theory has also given rise to the
possibility that the standard model gauge fields arise from
branes rather than in the bulk of space-time. This corresponds to
novel compactifications in which gravity lives in
the bulk of the ten dimensional space-time, but the other
observed fields (quarks, leptons, gauge particles etc.) live on a
brane of lower dimension\cite{SEIBWIT}. In particular we might
imagine a scenario in which the standard model fields live on a
three brane, the directions transverse to the three brane
being compact, and the directions tangential to the three
brane describing the usual three dimensional space. There may
also be other branes, separated from us in the extra directions,
forming `shadow worlds'! This situation has been illustrated in
Fig.\ref{f6}.  Since the
usual gauge and matter fields live on the brane, they do not see
the extra transverse directions; the only long range
interaction which sees the extra
directions is gravity. 
This allows the extra dimensions to be
much larger ($\sim 1mm$) than in the conventional
compactification scheme, the most stringent bound coming from
tests involving the inverse square law of Newtonian
gravity.\cite{KAPLUN} This fact has been used in recent proposals
for superstring model building.\cite{DIM}

\section{Summary}

It is now time to summarise the main results. We have seen that
string theory has had a reasonable success in providing a
consistent quantum theory of gravity. In particular we have
achieved:
\begin{itemize}
\item Finiteness of perturbation theory,
\item Partial resolution of the problems associated with quantum
theory of black holes, 
\item Explicit realization of holographic principle for a special
class of space-time, and
\item (3+1) dimensional theories with gravity, gauge
interactions, chiral fermions and $\NN=1$ supersymmetry, closely
resembling the standard model.
\end{itemize}
String theory has also provided us with an internally consistent
and beautiful theory. In particular string duality provides us
with
\begin{itemize}
\item Unification of many different string theories,
\item Unification of elementary and composite particles, and
\item Unification of classical and quantum effects.
\end{itemize}
Progress in string theory has also dramatically improved our
understanding of supersymmetric quantum field theories.

Unfortunately, there are still no concrete new predictions of string
theory at low energies. We shall have to wait and see if the
situation improves during the next few years.

\end{document}